\documentstyle[psbox]{WORKSHOP}
\newcommand{\asca}{{\it ASCA} }

\newcommand{\rxte}{{\it RXTE} }

\begin{document}

% TITLE OF THE PAPER
%  If the title is too long for a single line, you can split it 
%  by putting two backslashes. 
%  You might want to put the subtitle. Then it should be inserted 
%  within {\large\sf  }.
%  e.g.:  
%     \title{ Too Long Title \\ for one line \\
%     {\large\sf Subtitle} }
\title{
Five-year Monitorings of TeV Blazars with $ASCA$ and $RXTE$\\ 
%{\large\sf  -- Brief Instructions for Users of the `WORKSHOP' Style
%File --} 
}

% AUTHOR(S) 
\author{
J.Kataoka,$^1$ T.Takahashi,$^2$ F.Takahara,$^3$ P.G.Edwards,$^2$ 
K.Hayashida,$^3$ S.Inoue,$^4$\\ 
N.Iyomoto,$^2$ N.Kawai,$^1$ G.M.Madejski,$^5$ C.Tanihata,$^2$ 
and S.J.Wagner,$^6$  
\\[12pt]  % TO BE SPACED WITH ONE LINE
%
% INSTITUTES OF AUTHORS
$^1$  Department of Physics, Faculty of Science, Tokyo
Institute of Technology, Tokyo, Japan\\
$^2$  Institute of Space and Astronautical Science, Kanagawa 229-8510, Japan\\
$^3$  Department of Earth and Space Science, Osaka University, Osaka, Japan\\
$^4$  Theoretical Astrophysics Division, National Astronomical
      Observatory, Tokyo, Japan\\
$^5$  Stanford Linear Accelerator Center, California, USA\\
$^6$  Landessternwarte Heidelberg, K\H{o}nigstuhl, Heidelberg, Germany\\
%
% please put the first author's initial and e-mail address below
{\it E-mail(JK): kataoka@hp.phys.titech.ac.jp} 
%            \_ Initial      \
%                             \_ E-mail address
}

\abst{
We study the temporal/spectral variability of two extragalactic TeV sources, 
Mrk 421 and Mrk 501, based on 5-year observations with the \asca and \rxte 
satellites. We found that the peak of the synchrotron emission exists just 
in the X-ray band and its position shifted from lower to higher energy when 
the source became brighter. Relationship between the peak energy  and peak 
luminosity showed quite different behavior in the two sources; Mrk 421 
showed very little change in the peak position (0.5$-$2 keV), while Mrk 501 
revealed the largest shift ever observed in blazars (1$-$100 keV).  
We analyze these X-ray data with the flux changes in TeV band, which are 
obtained from 35  $truly$ simultateneous observations.  
Very different spectral evolution of both objects indicates some 
differences in the electron acceleration mechanism at work during the 
flares. We argue that the flux variability of Mrk 421 is associated with 
an increase in the number of electrons, while the flare of Mrk 501 is 
mostly due to the large changes in maximum energy of electrons. 
We also discuss the characteristic temporal variability of TeV sources,
and implications for the X-ray emitting site in the relativistic jet.
}
\kword{blazars: time variability --- spectrum: evolution}

\maketitle
\thispagestyle{empty}

\section{Introduction}

Blazars, a subclass of AGNs, have outstanding properties in several aspects.
They exhibit the most rapid and the largest amplitude variations of all 
AGNs.  Recent observations with the EGRET 
instrument on-board the 
{\sl Compton Gamma-Ray Observatory} ({\sl CGRO}) reveal that more than 60 
AGNs are bright $\gamma$--ray emitters (e.g., Hartman et al. 1999).
Remarkably, most of AGNs detected by EGRET were classified as blazars. 
Observations with ground-based Cherenkov telescopes further confirmed 
$\gamma$-ray emission extending up to TeV energies for a number of
nearby blazars.

Overall spectra of blazars (plotted as $\nu$ $F_{\nu}$) have at least
two pronounced continuum components: one peaking between IR and X--rays, 
and another in the $\gamma$--ray regime (e.g., von Montigny et al. 1995). 
The strong polarization observed in the radio and optical bands 
(Angel \& Stockman 1980) implies that the lower energy component is 
produced by synchrotron radiation of relativistic electrons in
magnetic fields, while inverse-Compton scattering by the same electrons
is dominant process responsible for the high energy 
$\gamma$--ray emission (Ulrich, Maraschi, \& Urry 1997). This
non-thermal radiation is thought to be emitted in a relativistic jet
pointing close to our line of sight (e.g., Urry \& Padovani 1995).

Multi-frequency spectra of blazars indicate their variety.
There are indications that particles are accelerated more efficiently 
in blazars having lower luminosities (e.g., Ghisellini et al. 1998).  
In fact, the luminosity of TeV blazars is the $lowest$ end of EGRET
detected blazars, while the emission peaks are
located at the $highest$ frequencies. In low luminosity objects, energy
loss by radiation would not be effective because there are less 
ambient photons to be Compton scattered to higher energies.  
Thus the TeV blazars gives an important opportunity to study the particle 
acceleration in blazar jets.

Mrk 421 and Mrk 501 are the proto-type of TeV $\gamma$-ray emitting blazars.
The first multi-frequency campaign of Mrk 421 including TeV energies was 
conducted in 1994 (Macomb et al. 1995; Takahashi et al. 1996). The 
contemporaneous observations implied correlated variability between the 
keV X--ray and TeV $\gamma$--ray emission, while the GeV flux and the radio 
to  UV fluxes showed less variability. The results from this campaign
are important because it suggested the possibility that a single
electron population is responsible for both the X--rays and TeV 
$\gamma$--rays, qualitatively agreeing with the Synchrotron self-Compton 
(SSC) scenario. However, the fact that the X-ray and TeV observations
were separated by one day, could introduce uncertainties to draw
concrete pictures. Simultaneous, uninterrupted observations are crucial 
to understand the blazar phenomenon.

In this paper we intensively study Mrk 421 and Mrk 501, with an emphasis 
on the observations with the X-ray satellites \asca and \rxte.  
Unlike most of the previous work which focused on the photon spectral 
properties and/or classification of blazars, we also study the rapid
time variability and spectral evolution in blazars.  The time
variability gives us important constraints on the dynamics operating in 
blazar jets, such as the acceleration, radiative cooling and particle escape. 
Taking the VLBI observations of superluminal motion into account, 
we discuss the internal jet structure and emission site in the jet.

\section{X-ray Observations and Data Analysis}
\subsection{Observations}

Mrk 421 and Mrk 501 were observed a number of times with the 
X-ray satellites \asca and/or \rxte. \asca observed Mrk~421 five times
with a net exposure of 546~ksec between 1993 and 1998.  In the 1998 
observation, the source was in a very active state and was detected at
its highest-ever level (Takahashi et al.\ 2000). \rxte observed Mrk~501
more than 100 times with a net exposure of 700~ksec between 1996 and
1998.  Mrk~501 was in a historical high-state in 1997 (Catanese et al.\
1997; Pian et al.\ 1998; Lamer \& Wagner 1999). Multiwavelength
campaigns, including a number of ToO (Target of Oppotunity)
observations, were conducted during this high state.
Observation logs are given in Table 1.

\begin{table}[h]
\caption{Observation log of Mrk 421 and Mrk 501}
\begin{center}
\begin{tabular}{lcccc} \hline\hline\\[-6pt]
            &          &                   & Exp.  &ID\\
Source      &Satellite & Obs. Time(UT)     &(ksec) &  \\
\hline
Mrk~421        & \asca &93.05.10 $-$ 05.11 &  43  & (a)\\
               & \asca &94.05.16 $-$ 05.17 &  39  & (b)\\
               & \asca &95.04.25 $-$ 05.08 &  91  & (c)\\
               & \asca &97.04.29 $-$ 05.06 &  70  & (d)\\
               & \asca &98.04.23 $-$ 04.30 &  280 & (e)\\
Mrk~501        & \rxte &97.04.03 $-$ 04.16 &  36  & (f)\\
               & \rxte &97.05.02 $-$ 05.15 &  51  & (g)\\
               & \rxte &97.07.11 $-$ 07.16 &  38  & (h)\\
               & \rxte &98.02.25 $-$ 10.02 & 568  & (i)\\
\hline
\end{tabular}
\end{center}
\end{table}

\subsection{X-ray Time Variability}

Figure~1 shows the long-term variation of fluxes, with both \asca and \rxte 
observations plotted. \asca observations of Mrk~421 spanned more than 
5~years (from 1993 to 1998), and show that the source exhibits variability by 
more than an order of magnitude. Blow-ups of the light curves taken in 
1995 and 1997 are given in the lower panel (Figure~1 (c),(d)). 
For Mrk~501,  we plot the \rxte data because the observations were
conducted much more frequently than the \asca observations. 
The \rxte observations spanned more than 3~years and significant flux
changes are clearly detected. Blow-ups of light curves are given in
Figure~1 (f)$-$(h). 

\begin{figure*}[t]
\centering
%\psbox[xsize=0.4#1,ysize=0.2#1,rotate=r]
\psbox[xsize=8.5cm]
{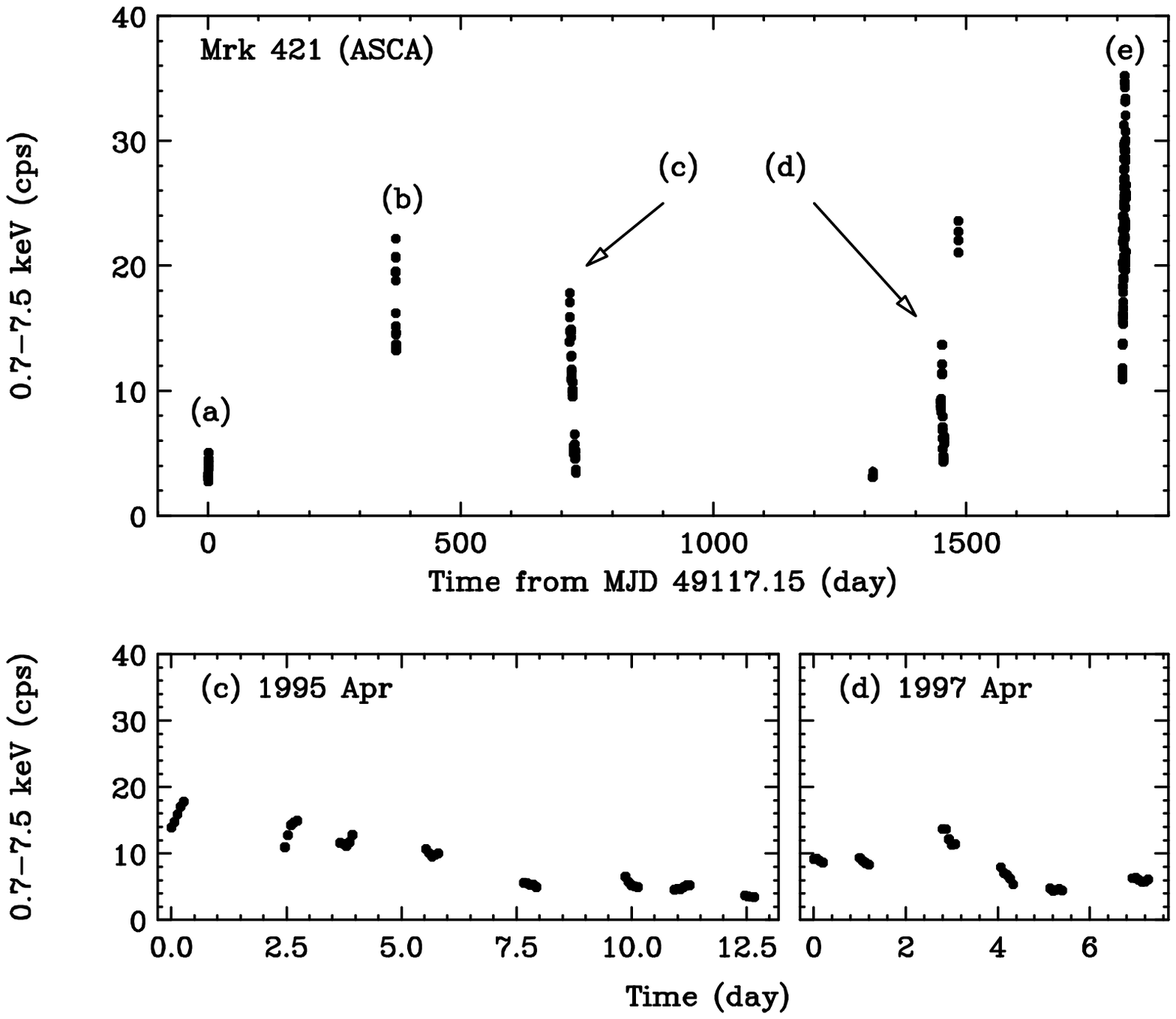}
\psbox[xsize=8.7cm]
{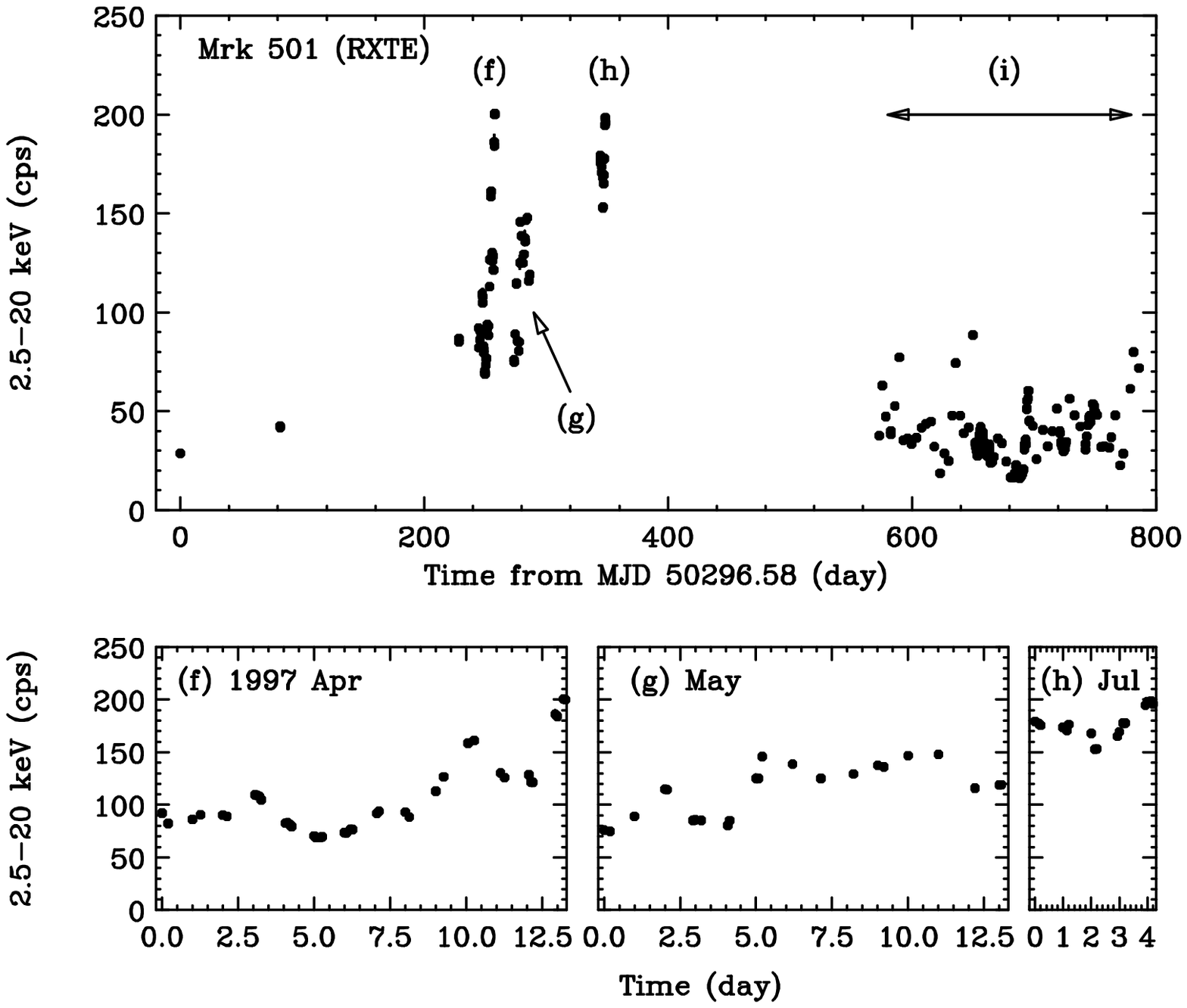}
\caption{Long-term flux variation of Mrk 421 ($left$) and Mrk 501 ($right$)}
\end{figure*}

To evaluate variability features in blazars, we examine the use of a 
numerical technique called the structure function (hereafter, SF).
While in theory the SF is completely equivalent to traditional Fourier 
analysis methods, it has several significant advantages. Firstly, it is 
much easier to calculate. Secondly, the SF is less affected by gaps in
the light curves.  The definitions of SFs and their properties are 
given by Simonetti et al.\ (1985). The first order SF is defined as
\begin{equation}
{\rm SF}(\tau) = \frac{1}{N}\sum[a(t) - a(t+\tau)]^2, 
\end{equation}
where $a(t)$ is a point of the time series (light curves) $\{$$a$$\}$ and 
the summation is made over all pairs separated in time by $\tau$. 
$N$ is the number of such pairs. 

\begin{figure*}[t]
\centering
%\psbox[xsize=0.4#1,ysize=0.2#1,rotate=r]
\psbox[xsize=8.0cm]
{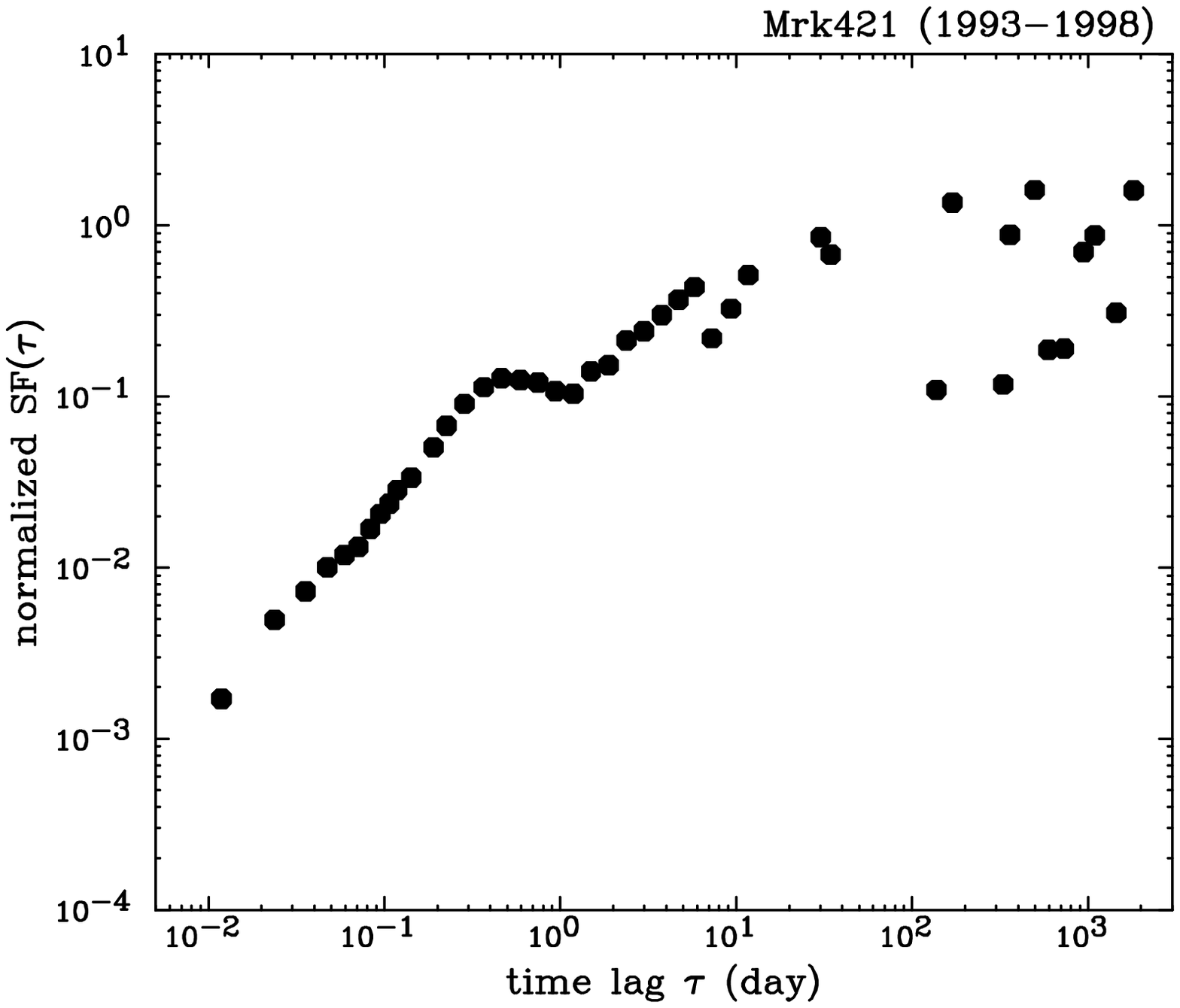}
\psbox[xsize=8.0cm]
{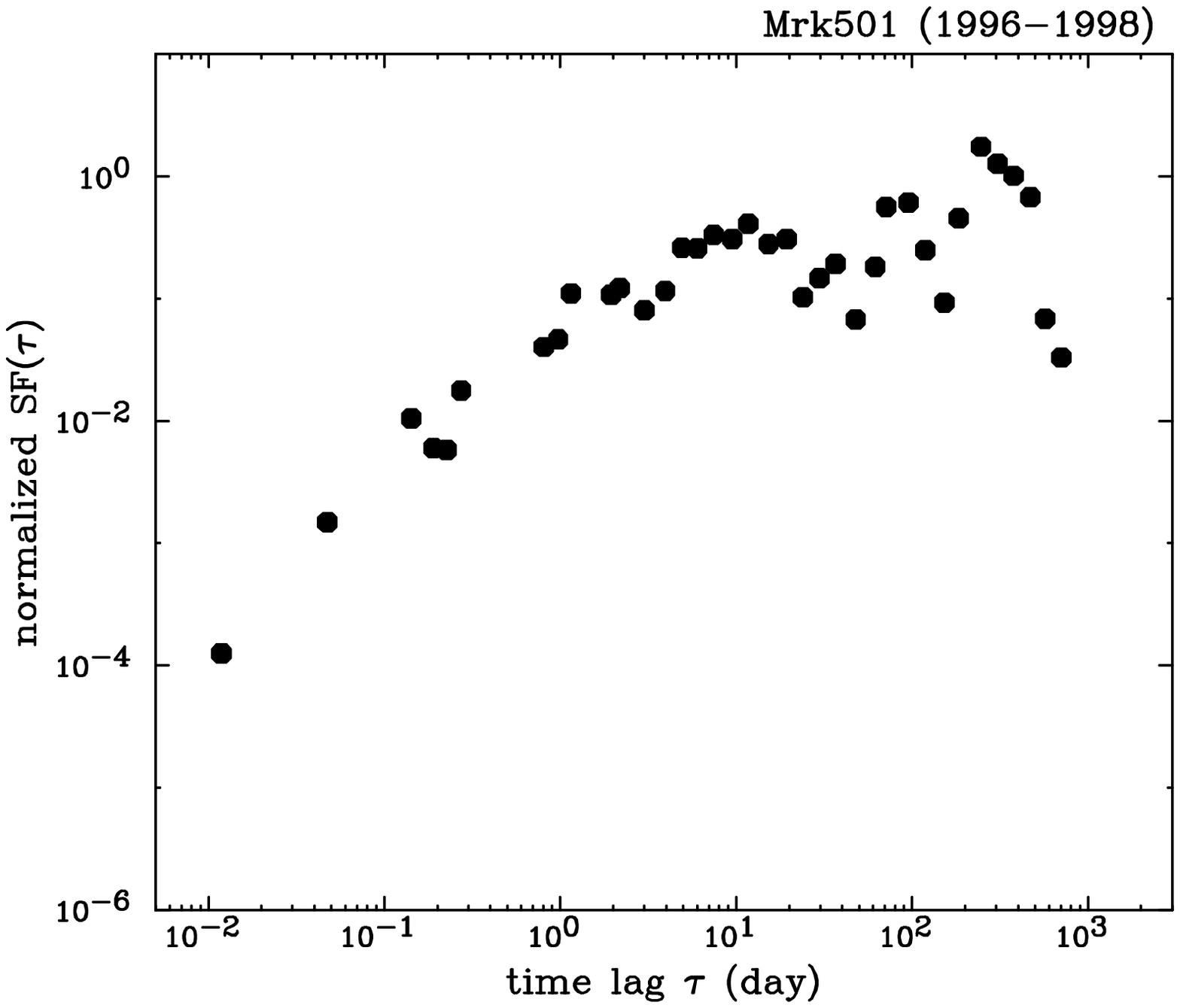}
\caption{Structure functions of Mrk 421 ($left$) and Mrk 501 ($right$)}
\end{figure*}

The structure functions calculated from the light curves (Figure~1) 
are given in Figure~2.  The SFs are normalized by the square of the mean 
fluxes, and are binned at logarithmically equal intervals. 
Both the SFs show a rapid increase up to $\tau$/day $\simeq$ 1, then  
gradually flatten to the observed longest time-scale of
$\tau$/day\,$\ge$\,1000. Fluctuations at large $\tau$
($\tau$/day\,$\ge$\,10) are due to the extremely sparse sampling of
data. In order to demonstrate the uncertainties caused by such sparse 
sampling, and to firmly establish the reality of the ``roll-over'', we 
simulated the long-term light curves following the $forward$ $method$ 
described in Iyomoto (1999). We found that (1) the PSD of the TeV sources
have at least one roll-over at 10$^{-6}$ Hz $\le$ $f_{\rm br}$ 
$\le$ 10$^{-5}$ Hz (1 $\le$ $\tau$/day $\le$ 10),  and (2) the PSD
changes its slope from $\propto$ $f^{-1 \sim -2}$ ($f$ $<$ $f_{\rm br}$) 
to $\propto$ $f^{-2 \sim -3}$ ($f$ $>$ $f_{\rm br}$) around the
roll-over (Kataoka et al. 2001a).

\subsection{X-ray Spectral Evolution}

\begin{figure*}[t]
\centering
%\psbox[xsize=0.4#1,ysize=0.2#1,rotate=r]
\psbox[xsize=7.2cm]
{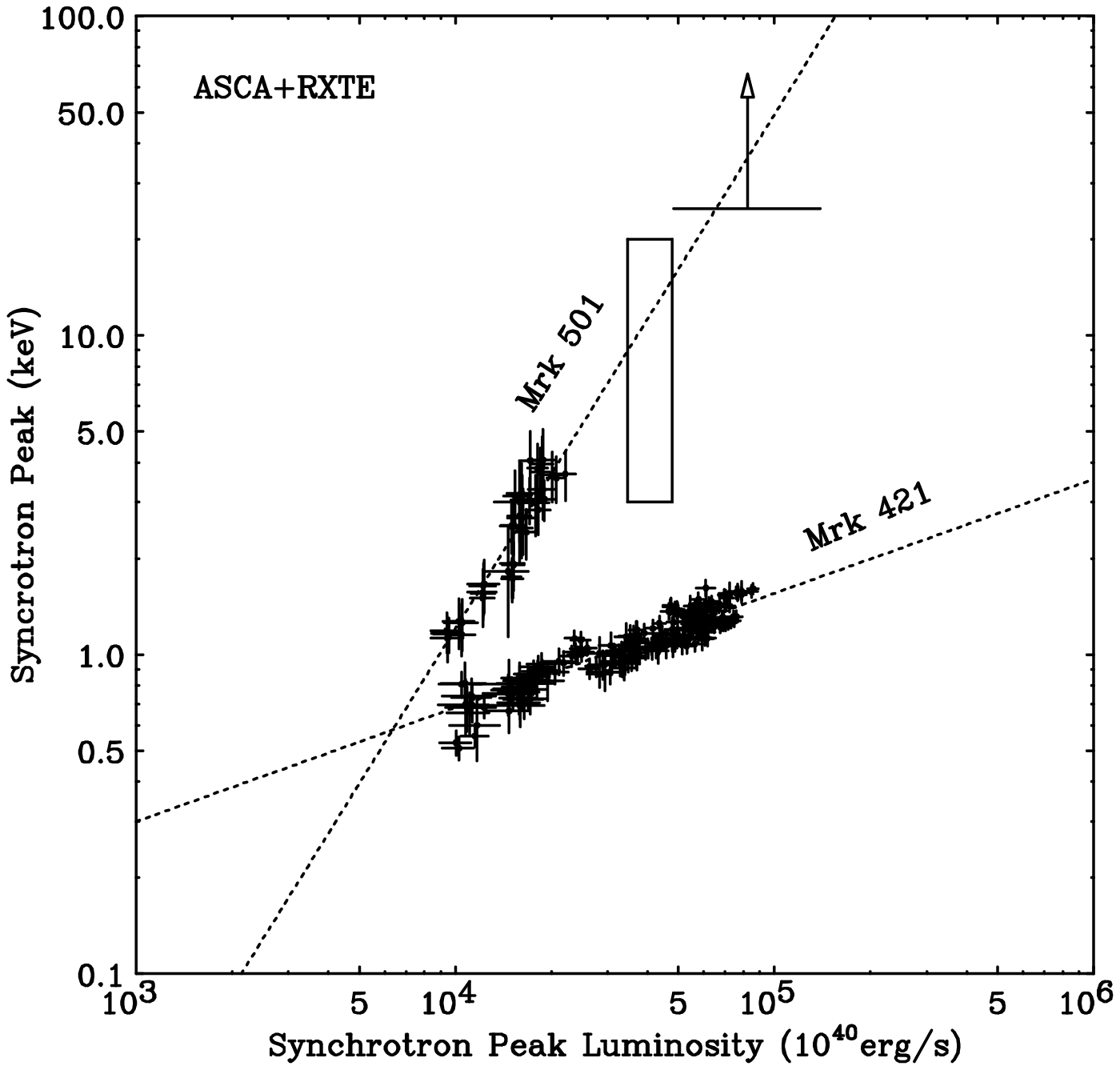}
\caption{Distribution of peak luminosity $L_{\rm p}$ versus peak energy
 $E_{\rm p}$}
\end{figure*}

We next performed model fitting to evaluate the X-ray photon spectra of 
Mrk 421 and Mrk 501. In general, the photon spectra of TeV sources show convex
shape, where the model with a single power law functions plus absorption
arising from neutral material could not fit the observed
spectra. We thus use another model which is expected to fit the spectrum
\begin{equation}
\frac{dN}{dE} = N_0 \times {\rm exp}(-N_{\rm H}^{\rm Gal} \sigma(E)) 
E^{-\Gamma}\times{\rm exp}( - E_{\rm c}/E),
\end{equation}
where $\Gamma$ is the photon index and  $E_{\rm c}$ is the cutoff energy, 
respectively. Using this relatively simple model, one can differentiate
the function analytically. The peak energy $E_{\rm p}$ in $\nu$ 
$F_{\nu}$ are given $E_{\rm p}$ = $\frac{E_{\rm c}}{\Gamma - 2}$. 
We cannot determine $E_{\rm p}$ for $\Gamma$ $\le$ 2, because in such
cases, the spectra are monotonously rising in the total energy band. 

We divided the observations into 5 ksec exporsure segments and fitted
the photon spectra for individual
segments. In this case, statistically acceptable fits were obtained.
The distribution of peak luminosity $L_{\rm p}$ versus peak energy $E_{\rm p}$ 
is summarized in Figure~3. The peak luminosity is simply calculated from
the luminosity at 
$E$ = $E_{\rm p}$. Mrk 421 shows very modest shifts in the peak position.
A clear correlation between the peak energy $E_{\rm p}$ and the 
peak luminosity $L_{\rm p}$ was detected for the first time for both 
Mrk 421 and Mrk 501: $E_{\rm p}$ $\propto$ $L_{\rm p}^{0.4}$ for Mrk
421 and $E_{\rm p}$ $\propto$ $L_{\rm p}^{1.6}$ for the case of Mrk 501.
Extrapolating the relation between $E_{\rm p}$ and $L_{\rm p}$, we
expect that synchrotron peak reaches to $\sim$ 100 keV in the high state 
of Mrk 501, which is the largest shift ever observed in blazars (Pian et
al. 1998; Kataoka et al. 1999).

\section{Multi-frequency Analysis}

In previous sections, we have investigated the rapid variability and 
spectral evolution of TeV blazars, highlighting the X-ray 
observations by \asca and \rxte. 
However, the overall spectral energy distribution of blazars generally
ranges over a very wide range -- from 
radio to TeV $\gamma$-ray bands -- and such spectra are 
one of the most important features allowing us to understand 
these sources. To reveal the multi-frequency properties of TeV blazars, we 
have to carry out multi-frequency monitoring campaigns 
of the source in various states of activity.  

Results from previous campaigns strongly suggest correlated variations  
in X-ray and TeV $\gamma$-ray fluxes, while variability is much less
pronounced 
in other energy bands. 
However, a problem remains that most of the data are taken 
either $non$-simultaneously or are very sparsely sampled 
(e.g., Macomb et al. 1995; Buckley et al. 1996).
The discussion based on those $quasi$-simultaneous data may be
incomplete. We need $exactly$ simultaneous 
monitoring, especially in the X-ray and TeV energy bands, 
to correctly understand the sources.  In the following, we summarize the 
results from $truly$ simultaneous campaigns of TeV blazars conducted 
from 1996 to 1998.

\begin{figure*}[t]
\centering
%\psbox[xsize=0.4#1,ysize=0.2#1,rotate=r]
\psbox[xsize=8.5cm]
{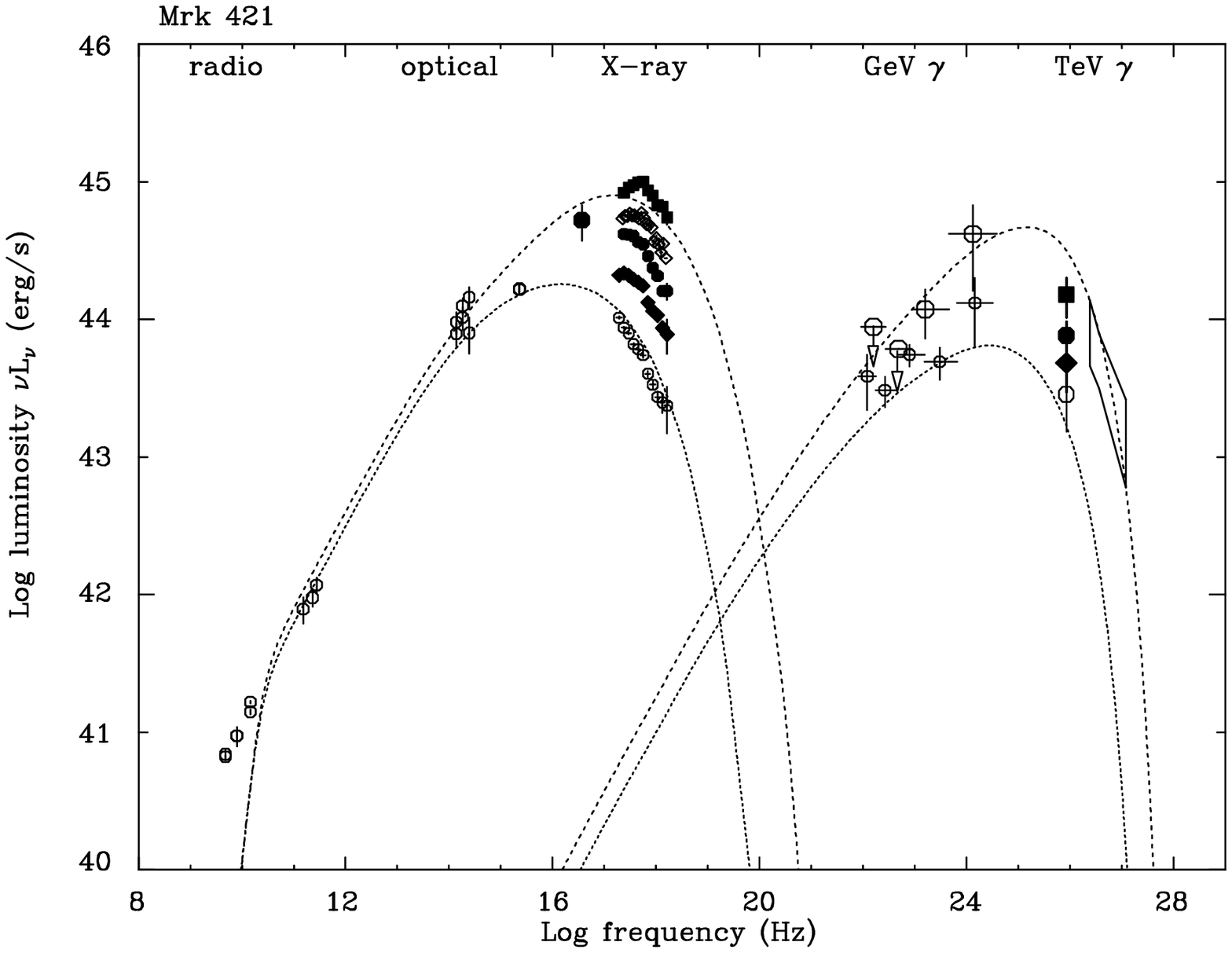}
\psbox[xsize=8.5cm]
{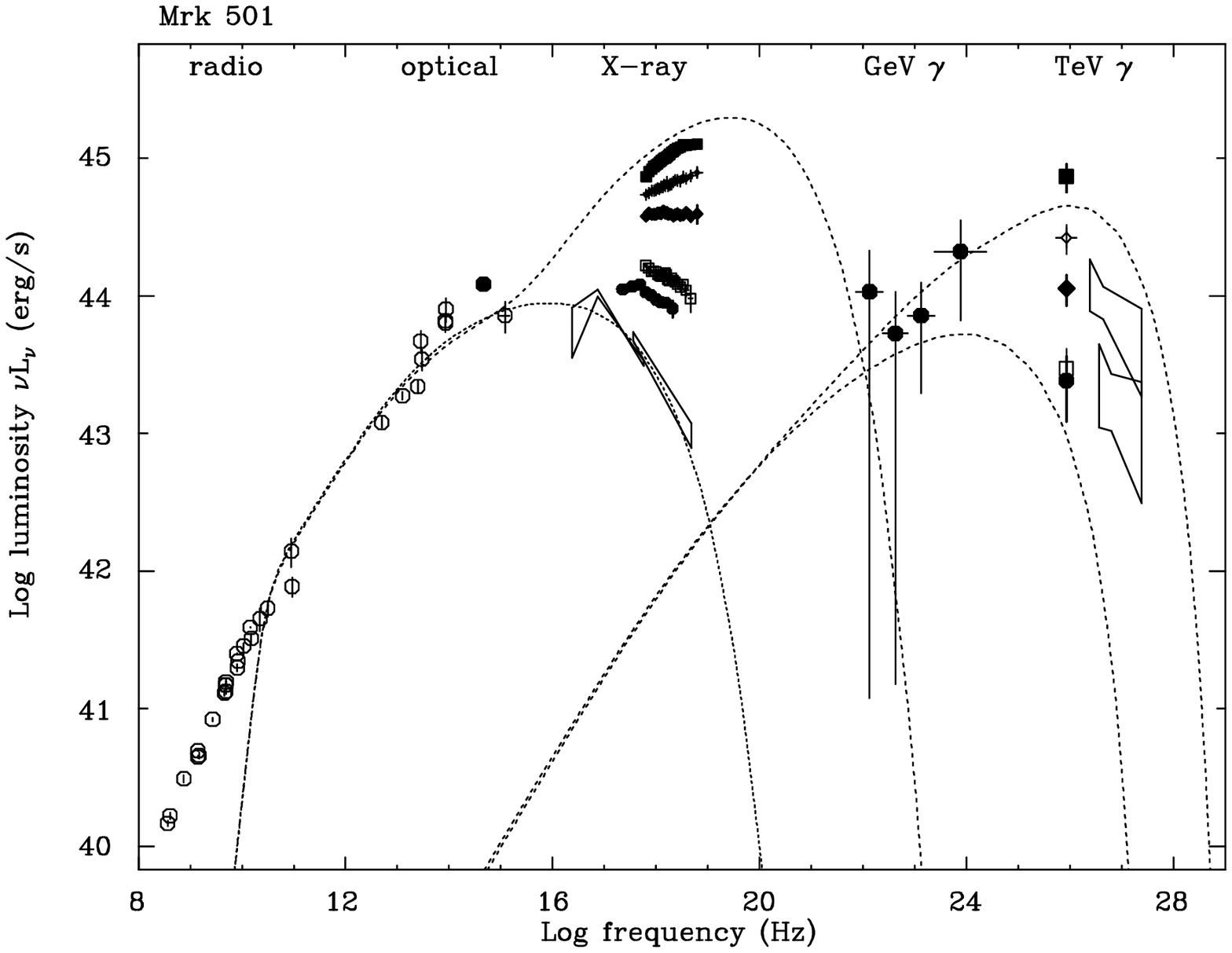}
\caption{Evolutions of multi-frequency spectra : Mrk 421 ($left$) and
 Mrk 501 ($right$). Dotted line represents the best fit SSC model for 
the quiescent/flare state. Figure from Kataoka et al. (2001b).}
\end{figure*}

Evolution of multi-frequency spectrum of Mrk 421 is shown in 
Figure 4($left$). Open circles come from Macomb et al 
(1995), while other symbols represent our 
new results taken exactly simultaneously in X-ray (also $EUVE$) and 
TeV $\gamma$-ray bands. Several important features are seen in the figure. 
First, a clear correlation can be seen between the keV X-ray flux and the
TeV $\gamma$-ray fluxes. The amplitude of TeV flux variation is less or almost 
comparable with that in the X-ray energy bands. Detailed analysis 
showed that the correlation is expressed as  [TeV $\gamma$-ray flux] $\propto$ 
[X-ray flux]$^{0.92 \pm 0.12}$. Second, the slope of X-ray photon
spectra ($\Gamma_{\rm X-ray}$ $\sim$ 3) are steep, and very similar to
those in the TeV energy band ($\Gamma_{\rm TeV}$ $\sim$ 3; Aharonian et
al. 1999b). Third, although X-ray flux changed dramatically in various 
seasons, only small changes are implied in the synchrotron peak
position, as we have quantitatively discussed in $\S$ 2.3.  

Evolution of the multi-frequency spectrum of Mrk 501 is shown in 
Figure 4 ($right$). Open circles are data from public 
archive (taken from Kataoka et al. 1999), while other symbols represent our 
new results derived from simultaneous monitoring in the X-ray and 
TeV $\gamma$-ray bands. For the data obtained in March 1996, 
we also plot the EGRET and optical data.  Compared to the results for 
Mrk 421, some significant differences are implied in the multiband specta. 
First, the changes in X-ray flux are accompanied by large shift in 
the position of the synchrotron peak. Second, the slope of X-ray spectra 
($\Gamma$) varies widely, ranging from $\Gamma_{\rm X-ray}$ $\sim$ 
1.7 to 2.5, while the photon index in the TeV energy band is almost 
unchanged; $\Gamma_{\rm TeV}$ $\sim$ 2.5 (e.g, Aharonian et al. 1999a). 
Third, amplitude of TeV flux variation is much larger than that in the X-ray
energy band. The correlation between two energy bands is expressed as  
[TeV $\gamma$-ray flux] $\propto$ [X-ray flux]$^{1.96 \pm 0.07}$. 

\section{Discussion: what can we learn?}

As first pointed out by Kataoka (2000), the little power of rapid 
variability ($\le$~1~day) in TeV sources provides important clues to the 
X-ray emitting site in the jet. The characteristic time-scale of each 
flare event $t_{\rm var}$ $\ge$ 1 day should reflect the size of the emission 
region, which we infer to be $\ge$ 10$^{16}$ ($\Gamma$/10)~cm in 
the source co-moving frame, if emitting blobs are approaching with Lorentz 
factors $\Gamma$ ($\Gamma$ $\simeq$ 10; Vermeulen \& Cohen 1994). 
This range of Lorentz factors is independently inferred from 
constraints that can be derived from the spectral shape and variability of 
TeV blazars (e.g., Tavecchio, Maraschi and Ghisellini 1998; Kataoka et al.\ 
1999). 

If the jet is collimated to within a cone of constant opening angle 
$\theta$ $\simeq$ 1/$\Gamma$ and the line-of-sight extent of shock 
is comparable with the angular extent of the jet,  one expects that the 
X-ray emission site is located at distances $D$ $\ge$ 10$^{17}$ 
($\Gamma$/10)$^{2}$ cm from the base of the jet. 
Only little variability shorter than $t_{\rm var}$ strongly suggests
that no significant X-ray emission can occur in regions closer than this
to the black hole. The relativistic electrons responsible 
for the X-ray emission are most likely accelerated and injected at shock 
fronts occurring in the jet (e.g., Inoue \& Takahara 1996). The lack of 
short term variability may then imply that shocks are nearly absent 
until distances of $D$ $\ge$ 10$^{17}$ ($\Gamma$/10)$^{2}$ cm. Further 
discussion of the characteristic timescale is given in Kataoka et
al. (2001a), which also suggests a possible link between the jet and the 
central engine.

In $\S$ 2.3, we found that the position of the synchrotron peak shifts
from lower to higher energy when the source becomes brighter. 
The difference of spectral evolution in Mrk 421 and Mrk 501 implies that 
quite different mechanisms were at work when the sources went into the 
flaring states. Since the peak luminosity is proportional to the number
of photons at peak ($n_{\rm ph}$($E_{\rm p}$)) multiplied by the peak energy, 
a simple relation can be found:  
\begin{equation}
L_{\rm p} \propto  E_{\rm p} n_{\rm ph}(E_{\rm p}) \propto 
\gamma_{\rm p}^2 n_{\rm e}(\gamma_{\rm p}), 
\end{equation}
where $\gamma_{\rm p}$ is the Lorentz factor of electron which emits photon of 
energy $E_{\rm p}$ and $n_{\rm e}(\gamma_{\rm p})$ is the number of electrons 
at $\gamma_{\rm p}$. We thus find the relations,
\begin{equation}
n_{\rm e}(\gamma_{\rm p}) \propto \gamma_{\rm p}^{3.6}\hspace{5mm}({\rm Mrk} 421),
\end{equation}
\begin{equation}
n_{\rm e}(\gamma_{\rm p}) \propto \gamma_{\rm p}^{-0.8}\hspace{5mm}({\rm Mrk} 501). 
\end{equation}
This implies that for the case of Mrk 421, increase of $\gamma_{\rm p}$ by 
a factor of 2 requires more than factor 10 $increase$ in number of electrons 
at peak energy. On the other hand, the same amount of increase of  
$\gamma_{\rm p}$ requires $decreases$ in number of electrons by a 
factor of 2 for the case of Mrk 501. 
Such variations in spectral behaviors may be associated with 
the difference of physical conditions in relativistic jets, which is 
discussed below. Full descriptions are given in Kataoka (2000) and 
Kataoka et al. (2001b). 

Assume that a substantial amount of gas (e.g., in a form of clouds) is
distributed in the jet. In these clouds, density of low-energy electrons is
enhanced as compared to the ambient, 
but other physical quantities such as magnetic field strength are unchanged. 
Those clouds essentially provide a plentiful source of low-energy electrons 
for the shock front. When a shock front overruns one of such 
clouds, fresh electrons are successively injected and assumed to undergo 
continuous acceleration by repeatedly crossing and recrossing the shock front, 
as well as simultaneously cooling by synchrotron radiation 
(e.g., Kirk, Rieger \& Mastichiadis 1998).  
In this scenario, number of electrons increases  significantly, but only 
small changes are 
implied for the maximum Lorentz factor as were observed in Mrk 421.

On the other hand, when clouds are absent or very sparsely 
distributed in the jet, flares may be produced in several diffent manners. 
For example, if the shock overruns the enhanced tangled-magnetic field 
region, this may cause changes in acceleration time of electrons, and hence 
increase the maximum Lorentz factor. Importantly, 
regardless of detailed models 
for flaring behaviour, total number of electrons is $conserved$ in 
this case. During the flare, acceleration can be assumed to be more efficient 
than radiative cooling, thus the present electron population as a whole 
will be accelerated to higher energies, but no additional electrons are 
supplied into the shock.

In latter case, numbers of electrons at the peak ($n_{\rm e}$ $(\gamma_{\rm p}$)) will $decrease$, reflecting the power-law 
shape of an electron population.  If the differential number density 
of electrons is expressed as $N_{\rm e}$ $\propto$ $\gamma^{-2}$ 
(standard shock), electron number decreases as 
$n_{\rm e}$($\gamma_{\rm p}$) 
$\propto$ $\gamma_{\rm p}$$N_{\rm e}(\gamma_{\rm p})$ 
$\propto$ $\gamma_{\rm p}^{-1}$.
Importantly, this relation is very close to the case we have observed 
in Mrk 501.

Present discussion based on the X-ray spectral evolution suggests       
very important implications for internal jet structures. 
Only a small shift of synchrotron peak observed in Mrk 421 may be 
associated with electron clouds $filling$ the jet, while the jet of 
Mrk 501 seems to be relatively empty. During the flare of Mrk 421, 
kinetic power of the shock is equally distributed to large number of 
low-energy electrons newly injected into the shock, thus increasing the 
number of high energy electrons.  Large shifts of synchrotron peak observed 
in Mrk 501, on the other hand, is possible only when the internal jet is 
rather sparse and transparent to the shock propagation. Kinetic power of 
the shock is spent to $increase$ the energies of individual electrons 
and hence $number$-$conservative$.

It may thus be worthwhile to compare our X-ray implications to the VLBI 
results. Accurate measurements of changes in the parsec-scale jet structure 
imaged with VLBI provide constraints on the jet kinematics and geometry. 
When combined with estimates of the Doppler beaming factor (determined, for 
example, from the X-ray time variability), the apparent motion of the jet 
components can be used to constrain the Lorentz factor of the jet and the 
angle of the jet to the line-of-sight. Although VLBI observations do not yet 
have an enough resolution to image the region production of 
the X-ray/$\gamma$-rays ($\sim$ 0.01 pc), they provide the highest resolution 
structual information available, and can image the region immediately 
downstream.

Mrk 421 and Mrk 501 are observed in a space VLBI project using the
$HALCA$ satellite and 12 ground stations. 
It is interesting that the subparsec- and parsec-scale jets of Mrk 421
and Mrk 501 appear to be weak relative to those of other blazars 
(Marscher et al. 1999). Most importantly, superluminal motions have been 
detected only for Mrk 501 ($v$ = 6.7 $c$; Giovannini et al. 1998), 
while subluminal motions were  implied for Mrk 421 ($v$ $\simeq$ 0.3
$c$; Piner et al. 1999).  
By assuming that the bulk and pattern jet velocity are comparable, 
Giovannini et al. (1998) derive that the beaming factor of Mrk 501 as 
$\delta$ $\sim$ 1.3$-$5.6. For the case of Mrk 421, the results are 
consistent with $no$-$beaming$.

In any case, the estimated beaming factors for both sources are relatively low 
compared with the lower limits derived from time variabilities in other 
wavebands; for examples, Takahashi et al. (1996) derived $\delta$ $\ge$ 5 
for Mrk 421 and Kataoka et al. (1999) derived $\delta$ $\ge$ 6 for Mrk 
501 using the X-ray/TeV $\gamma$-ray time variability.  
Marscher (1999) disscussed that rather low superluminal apparent speeds 
and lackluster variability properties of the radio jets evidences that 
the bulk flow of the jets $decelerates$ from X-ray/TeV $\gamma$-ray emitting 
section ($\sim$ 0.01 pc) to the radio emitting region ($\sim$ 1 pc).
The weak subparsec- and parsec-scale jets of those objects are readily  
understood as the consequence of heavy energy and monmentum loss in the 
upstream of the jet where most of the energy and momtentum of the 
relativistic electrons are transfered to the radiation in the X-ray and TeV 
$\gamma$-rays. 

Remarkably, very different manners of X-ray spectral evolutions of Mrk 421 and 
Mrk 501 presented in this paper is exactly consistent with those 
VLBI observations. Our X-ray observations predict that relativisitc outflows 
of Mrk 421 will be decelerated faster than that for Mrk 501, because the jet 
of Mrk 421 is filled with low-energy materials and kinetic energy of 
outflows are more efficiently dissipated during the propagation. 
The absence of superluminal motion of  Mrk 421 in radio bands thus indicates 
that the high energy outflow has been sufficiently decelerated when it reaches 
to more distant, radio emitting region.

%\section{Conclusion}

\section*{References}
\re
Aharonian, F., et al. 1999a, A \& A, 349, 29
\re
Aharonian, F., et al. 1999b, A \& A, 350, 757
\re
Buckley, J. H., et al. 1996, ApJ, 472, L9
\re
Catanese, M., et al. 1997, ApJ, 487, L143
\re
Ghisellini, G., et al. 1998, MNRAS, 301, 451 
\re
Giovannini, G., et al. 1998, in the proceedings of ``BL Lac phenomenon'' 
Turku, Finland
\re
Hartman, R. C., et al. 1999, ApJS, 123, 79
\re
Inoue, S., \& Takahara, F. 1996, ApJ, 463, 555 
\re
Iyomoto, N., 1999, Ph.D Thesis, University of Tokyo
\re
Kataoka, J., et al. 1999, ApJ, 514, 138 
\re
Kataoka, J., 2000, Ph.D Thesis, University of Tokyo
\re
Kataoka, J., et al. 2001a, ApJ, in press (astro-ph/0105022) 
\re
Kataoka, J., et al. 2001b, submitted 
\re
Kirk, J. G., et al. 1998, A\&A, 333, 452
\re
Lamer, G., \& Wagner, S. J. 1998, A \& A, 331, L13
\re
Macomb, D. J., et al. 1995, ApJ, 449, L99
\re
Marscher, A. P., 1999, Astroparticle Physics, 11, 19
\re
Pian, E., et al. 1998, ApJ, 492, L17
\re
Piner, B, G., et al. 1999, ApJ, 525, 176
\re
Simonetti, J, H., et al.  1985, ApJ, 296, 46
\re
Takahashi, T., et al. 1996, ApJ, 470, L89
\re
Takahashi, T., et al. 2000, ApJ, 542, L105
\re
Tavecchio, F., et al. 1998, ApJ, 509, 608 
\re
Ulrich, M. --H., et al. 1997, ARAA, 35,445
\re
Urry, C. M., \& Padovani, P., 1995, PASP, 715, 803
\re
Vermeulen, R. C., \& Cohen, M. H. 1994, ApJ, 430, 467

\label{last}

\end{document}